\documentclass[runningheads]{llncs}
\usepackage{graphicx}
\usepackage[misc]{ifsym}
\usepackage{array}
\begin{document}
\title{Exploring 3D U-Net Training Configurations and Post-Processing Strategies for the MICCAI 2023 Kidney and Tumor Segmentation Challenge}
\author{Kwang-Hyun Uhm\inst{1} \and
Hyunjun Cho \inst{1} \and
Zhixin Xu\inst{1} \and
Seohoon Lim\inst{1} \and
Seung-Won Jung\inst{1} \and
Sung-Hoo Hong\inst{2} \and
Sung-Jea Ko \inst{1, 3} \textsuperscript{\Letter}  
} 
%index{Uhm, Kwang-Hyun}
%index{Xu, Zhixin}
%index{Lim, Seohoon}
%index{Jung, Seung-Won}
%index{Hong, Sung-Hoo}
%index{Ko, Sung-Jea}
%
\authorrunning{F. Author et al.}
\institute{Korea University \and
The Catholic University of Korea, Seoul  \and
MedAI\\
\email{khuhm@dali.korea.ac.kr}
}

\maketitle

\begin{abstract}
In 2023, it is estimated that 81,800 kidney cancer cases will be newly diagnosed, and 14,890 people will die from this cancer in the United States. Preoperative dynamic contrast-enhanced abdominal computed tomography (CT) is often used for detecting lesions. 
However, there exists inter-observer variability due to subtle differences in the imaging features of kidney and kidney tumors.
In this paper, we explore various 3D U-Net training configurations and effective post-processing strategies for accurate segmentation of kidneys, cysts, and kidney tumors in CT images.
We validated our model on the dataset of the 2023 Kidney and Kidney Tumor Segmentation (KiTS23) challenge.
Our method took the second place in the final ranking of KiTS23 challenge on unseen test data with an average Dice score of 0.820 and an average Surface Dice of 0.712.
\end{abstract}

\keywords{Kidney cancer   \and Medical image segmentation \and 3D U-Net.}

\section{Introduction}
In 2023, it is estimated that 81,800 kidney cancer cases will be newly diagnosed, and 14,890 people will die from this cancer in the United States~\cite{kidneycancer}.
Kidney cancer is one of the 10 most common cancers, and by far the most common type of kidney cancer is renal cell carcinoma (RCC), which occurs in 9 out of 10 cases of all kidney cancer~\cite{npj_uhm}.
Preoperative dynamic contrast-enhanced abdominal computed tomography (CT) is often used for the detection and evaluation of renal tumors~\cite{2022_uhm}.
However, there are some overlaps in image-level features between kidneys, cysts, and renal tumors, which make accurate segmentation difficult and cause inter-observer variation.
These clinical issues point to the need to develop automatic systems that can reduce misdiagnosis and inter-observer variation.

In this paper, we explore various 3D U-Net training configurations and effective post-processing strategies for accurate segmentation of kidneys, cysts, and kidney tumors in CT images.
We investigate a wide variety of training configurations including training at different scales, cascade training approaches, and region-based training.
We also introduce post-processing approaches which aim at improving the performance by effectively combining the predictions from the models trained in different training configurations.
We validated our model on the dataset of 2023 Kidney and Kidney Tumor Segmentation (KiTS23) challenge, which includes data from previous challenges, ~\textit{e.g.}, KiTS19~\cite{KiTS19Challenge}.

%###########################
%###########################
\section{Methods}
% This is a good place to briefly introduce your approach and provide a teaser figure that illustrates how it works and any ways in which it might be unique.

%-------------------------------------------------------------------------
\begin{figure}[!t]
\centering
\includegraphics[scale=1.0]{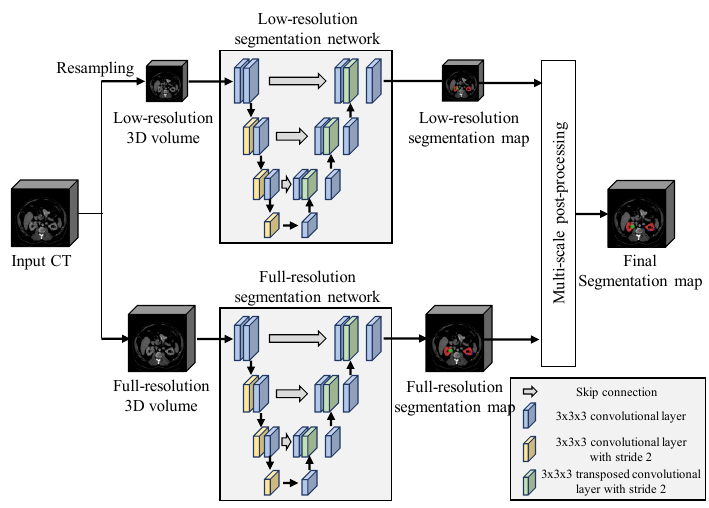}
\caption{Overview of our multi-scale prediction fusion framework. }
\label{fig:figure_main}
\end{figure}
%-------------------------------------------------------------------------

In this paper, we propose to perform segmentation at both low-resolution and full-resolution and then combine those two predictions based on a task-specific post-processing scheme, as shown in Fig.~\ref{fig:figure_main}. 
Low-resolution 3D CT volumes are generated by resampling original input CT images. 
Two independent 3D U-Nets are utilized to produce low-resolution segmentation maps and full-resolution segmentation maps from low-resolution 3D CT volume and full-resolution CT volume, respectively.
The final segmentation map is produced by multi-scale post-processing which takes the predictions of low- and full-resolution segmentation maps and combines them with domain-specific procedures.

%###########################
\subsection{Training and Validation Data}
Our submission made use of the official KiTS23 training set alone~\cite{KiTS19Data}.
We divide the provided data into a training set and a validation set at a ratio of 4:1.

\subsection{Preprocessing}
% Full description of your pre-processing strategy. Make sure to answer the following questions

% \begin{enumerate}
%     \item How was the data spatially transformed and/or resampled?
%     \item How were the HU values normalized/transformed, if at all?
%     \item How was the data cleaned and/or filtered, if at all?
% \end{enumerate}

We follow the way in the nnU-Net~\cite{nnunet} to preprocess the training data.
The spacing of all official CT images is the same on the x-axis and y-axis, but different on the z-axis.
The original training data have a median voxel spacing of 0.78 × 0.78 × 3.0 mm$^3$, and the median volume shape of 512 × 512 × 104 voxels.
For the training of low-resolution segmentation network, input data is resampled to have a spacing of 1.84 × 1.84 × 2.36 mm$^3$, which results in a median volume of 217 × 217 × 177 mm$^3$ voxels.
On the ohter hand, for the training of full-resolution segmentation network, input data is resampled to have a spacing of 0.78 × 0.78 × 1.0 mm$^3$, which results in a median volume of 512 × 512 × 417 mm$^3$ voxels.

We clip each case’s intensity values to the 0.5 and 99.5 percentiles of the intensity values in the foreground regions across the training set, \textit{i.e.}, the range of [-58, 302]. 
We subtract the mean value of 103 and then divide it by the standard
deviation of the intensities in the foreground regions, which is 73.3.
The foreground class oversampling is used to enforce more than a third of the samples in a batch contain at least one randomly chosen foreground class.
During training, patches with shape 128 × 128 × 128 are sampled and input
to the network. 
A variety of data augmentation techniques are applied on the fly during
training: rotations, scaling, mirroring, etc.

\subsection{Proposed Method}
% This will be your longest section. Use this space to describe your approach in detail, adding ``subsubsections'' as desired. Make sure cover the following topics:

% \begin{enumerate}
%     \item Network architecture, including hyperparameters such as strides, kernel sizes, etc.
%     \item Loss function, including any class-weighting or other hyperparameters
%     \item Optimization strategy, including any hyperparameters
%     \item Validation strategy, including selection criteria for final models 
%     \item Ensembling strategy, if any
%     \item Any other post-processing performed on the predictions to enhance the output
% \end{enumerate}
\subsubsection{Network Architecture}
We use 3D U-Net architecture~\cite{cicek2016unet3d} for both low-resolution and full-resolution segmentation networks.
The U-Net consists of an encoder and decoder, where for all convolutions in the network we use  3 × 3 × 3 convolution kernels.
Each block in the encoder consists of a sequence of two convolutional layers each of which followed by instance normalization and LeakyReLU activations. In the decoder, upsampling is done by 3 × 3 × 3 transposed convolutions.
At the last convolutional layer, it outputs the probability distributions for background, kidney, cyst, and tumor for each voxel.
There are a total of 6 stages for each encoder and decoder.
We use cross-entropy loss and dice loss for the training.
The stochastic gradient descent strategy is used for the optimization.
We investigate various network configurations.
For example, we increase all the channel numbers of convolutions in the network by two times.
This increases the model parameter by two times, and also the training time.
Another variant is a residual 3D U-Net which replaces plain convolution blocks in the encoder with residual blocks. 
Also, we tested the region-based training, which uses three sigmoid activations after the final convolutional layer to produce probabilities for each region, where regions are defined by "kidney and masses", "masses", and "tumor".
For the cascade model, the output of the low-resolution segmentation network is concatenated to the input image and then fed to the high-resolution segmentation network.

\subsubsection{Multi-Scale Post-Processing}
Based on our analysis of the results of the validation set, the low-resolution segmentation network produces well-localized segmentation parts but lacks sufficient details, while the full-resolution segmentation model provides finely detailed boundaries of kidneys and masses but generates some false positives around backgrounds.
Therefore, we remove segmented foreground blobs in full-resolution segmentation maps which do not belong to the foreground blobs in low-resolution segmentation maps.
Moreover, to reduce tumor false positives (FPs), we first perform the connected component analysis for the tumor class, and then we treat the tumor regions in each segmentation map as true positives if they have sufficient overlap with the tumor regions of the segmentation map from another scale. Specifically, the two tumor regions from low- and full-resolution segmentation maps should have a Dice coefficient greater than 0.3 to be determined as true tumor regions. 
Finally, we join the regions of predicted segmentation parts from both the full-resolution and low-resolution segmentation maps to complement each other.
This enables us to take advantage of both the low-resolution and high-resolution segmentation predictions, boosting the final segmentation performance.
In addition, we find the convex hulls of tumor blobs in the predicted segmentation maps and then merge the labels inside the detected hulls to reduce noisy prediction results.
We remove foreground blobs that have an area smaller than 10,000 mm$^3$.
%###########################
%###########################
%-------------------------------------------------------------------------
\begin{table}[t]
\caption{Results of experiments on the validation set.}
\begin{center}
\begin{tabular}{ m{4.5cm} m{2cm} m{2cm} m{2cm} m{1.2cm} }
\hline 
Method  & Kidney & Masses  & Tumor  & Average \\
\hline
Low-resolution   & 0.973  & \textbf{0.858} & 0.794 & 0.875 \\
Low-resolution - channel$\times$2   & 0.973  & 0.851 & 0.771 & 0.865 \\
Low-resolution - residual   & 0.973  & 0.856 & 0.794 & 0.874 \\
Full-resolution   & 0.977  & 0.840 & 0.790 & 0.869 \\
Full-resolution - channel$\times$2   & 0.975  & 0.840 & 0.790 & 0.868 \\
Full-resolution - residual   & \textbf{0.979}  & \textbf{0.858} & 0.803 & 0.880 \\
Full-resolution - batch 4   & 0.978 & 0.857 & 0.801 & 0.879 \\
Cascade   & \textbf{0.979}  & \textbf{0.858} & 0.804 & 0.880 \\
Region-based training   & 0.975  & 0.851 & 0.790 & 0.872 \\
Ours (w/ post-processing)  & \textbf{0.979} & 0.857 & \textbf{0.826} & \textbf{0.887} \\
\hline
\end{tabular}
\end{center}
\label{tab:results}
\end{table}

\section{Results}
% You will be given the opportunity to add your test set results to this section once they are announced, but before then, please use this section to describe how your approach appears to perform on the publicly available data alone. Be sure to include:

% \begin{enumerate}
%     \item Metric values during validation -- this is a good place for a table that breaks these metrics down by class
%     \item Some examples of predictions next to human-labels -- especially valuable if they illustrate a common mode of disagreement
% \end{enumerate}

% This is also a good place to include details about how training went, such as how long training and inference took on a given computing setup.

We validated our model on the dataset of the 2023 Kidney and Kidney Tumor Segmentation (KiTS23) challenge.
We report performances of baselines and our models evaluated on the validation set.
We report Dice scores for regions of "Kidney and Masses", "Kidney Masses", and "Kidney Tumor".   
For brevity, we denote in the tables in this section the dice scores as "Kidney", "Masses", and "Tumor".

We summarize the quantitative results in Tab 1. All the results are based on
the validation set, which contains 98 cases. We can see that our method outperforms other baselines by a large margin in terms of average Dice score. The average Dice is 0.887, and Dice
for kidney, kidney masses, and kidney tumors are respectively 0.979, 0.857, and
0.826. For the tumor segmentation, our algorithm performs significantly better than the baseline. These results demonstrate the effectiveness of our multi-scale post-processing strategy.

Our final submission involves three base networks including ``Low-resolution", ``Low-resolution - residual", and ``Full-resolution - batch 4" which show better performance than others in several folds. We apply our multi-scale post-processing strategy to two possible pairs of low- and full-resolution segmentation results and join the two post-processed results to make the final segmentation map.
%-------------------------------------------------------------------------
\begin{table}[t]
\caption{KiTS23 leaderboard for final results on test data (top-5).}
\begin{center}
\begin{tabular}{ c m{3.5cm} m{3.5cm} m{1.2cm} m{1.2cm} m{1.2cm}}
\hline 
Rank & Team & Affiliation  & Dice & Surface Dice  & Tumor Dice  \\
\hline
1 & Andriy Myronenko et al. & NVIDIA  & 0.835  & 0.723 & 0.756  \\
2 & Kwang-Hyun Uhm & Korea University  & 0.820  & 0.712 & 0.738  \\
3 & Yasmeen George & Monash University  & 0.819  & 0.707 & 0.713  \\
4 & Shuolin Liu &  Independent Researcher  & 0.805  & 0.706 & 0.697  \\
5 & George Stoica et al.  & University of Lasi, SenticLab   & 0.807  & 0.691 & 0.713  \\

\hline
\end{tabular}
\end{center}
\label{tab:results}
\end{table}

%###########################
%###########################
\section{Discussion and Conclusion}
In this paper, we explore various 3D U-Net training configurations and effective post-processing strategies for accurate segmentation of kidneys, cysts, and kidney tumors in CT images.
We investigate a wide variety of training configurations including training at different scales, cascade training approaches, and region-based training.
We also introduce post-processing approaches that aim at improving performance by effectively combining the predictions from the models trained in different training configurations.
We validated our model on the dataset of 2023 Kidney and Kidney Tumor Segmentation (KiTS23) challenge.

Our approach won second place in KiTS23 Challenge. On the test set, our final model
obtained Dice scores of 0.948, 0.776, and 0.738 and Surface Dice (SD) scores of 0.899, 0.635
and 0.602 for kidney, masses, and tumors, respectively. 
It is important to note that by carefully combining the results of 3D U-Nets trained from different scale settings, we can obtain much better performance than the results of individual models.
Specifically, we achieved a much higher tumor dice score than lower-ranked teams, which demonstrates that our multi-scale tumor prediction aggregation strategy is effective for accurate tumor segmentation.
Our method can be applied to other multi-scale approaches in current literature to improve the segmentation performance for accurate diagnosis.
Our work is only validated on KiTS23 Challenge dataset and lacks sufficient experimental validation on multiple datasets. 
In the future, we will expand the experiment settings of our method to different datasets for comprehensive evaluation of generality.  

\section*{Acknowledgment}
We thank the KiTS competition organizers, data providers, and annotators for
their great effort in the challenge. We further thank the creator
and contributors to the nnU-Net framework.

This work was supported by the Korea Medical Device Development Fund grant
funded by the Korea government (the Ministry of Science and ICT, the Ministry of Trade,
Industry and Energy, the Ministry of Health \& Welfare, the Ministry
of Food and Drug Safety) (Project Number: 1711195432, RS-2020-KD000096).
%!% I (Nicholas Heller) would like to personally acknowledge Jun Ma for sharing his FLARE21 short paper template with me, which I used as a starting point for this one.

\bibliographystyle{splncs04}
\bibliography{paper16}

\end{document}